\documentclass[aps,prb,superscriptaddress,twocolumn,showpacs]{revtex4}

\usepackage{amsmath}
\usepackage{bm}
\usepackage{graphicx,epsfig}

\renewcommand{\v}[1]{{\bf #1}}
\newcommand{\be}{\begin{equation}}
\newcommand{\ee}{\end{equation}}
\newcommand{\bd}{\begin{displaymath}}
\newcommand{\ed}{\end{displaymath}}
\newcommand{\ba}{\begin{eqnarray}}
\newcommand{\ea}{\end{eqnarray}}
\newcommand{\nn}{\nonumber \\}
\newcommand{\bpm}{\begin{pmatrix}}
\newcommand{\epm}{\end{pmatrix}}

\begin{document}

\title{Theory of magneto-electric susceptibility in multiferroic chiral magnets}

\author{Ye-Hua Liu}
\affiliation{Zhejiang Institute of Modern Physics and Department of Physics,   Zhejiang University, Hangzhou 310027, People's Republic of China}

\author{Jung Hoon Han}
\affiliation{Department of Physics, Sungkyunkwan University, Suwon 440-746,   Korea}
\affiliation{Asia Pacific Center for Theoretical Physics, Pohang, Gyeongbuk   790-784, Korea}

\author{A. A. Omrani}
\affiliation{Laboratory for Quantum Magnetism, Ecole Polytechnique F\'ed\'erale de Lausanne (EPFL), 1015 Lausanne, Switzerland}

\author{H. M. R{\o}nnow}
\affiliation{Laboratory for Quantum Magnetism, Ecole Polytechnique F\'ed\'erale de Lausanne (EPFL), 1015 Lausanne, Switzerland}

\author{You-Quan Li}
\email[Electronic address: ]{yqli@zju.edu.cn}
\affiliation{Zhejiang Institute of Modern Physics and Department of Physics,   Zhejiang University, Hangzhou 310027, People's Republic of China}

\date{\today}

\begin{abstract}
  We present a theoretical examination of the magneto-electric response in the recently discovered multiferroic insulator Cu$_2$OSeO$_3$. Combining Monte Carlo simulation and Ginzburg-Landau analysis we predict the response in each of the magnetic phases, including helical, conical, ferromagnetic and Skyrmion crystal phases, both for thin film and bulk systems. A common feature for all non-collinear phases is that the magneto-electric susceptibility increases linearly with the applied magnetic field. Being both calculable and measurable, the magneto-electric susceptibility can serve as a new powerful probe to detect and investigate magnetic phases and phase transitions in multiferroic chiral magnets in general.
\end{abstract}


\maketitle

\section{Introduction}

The discovery of nontrivial topological structures in condensed matter physics attracts significant attention. Starting from the prediction \cite{bogdanov1,bogdanov2,han1} and observation \cite{pfleiderer1,pfleiderer2,tokura,FeGe,MnGe} of the Skyrmion crystal phase in chiral magnets such as MnSi, the study of stabilization and manipulation of the Skyrmions has become an important subject in the field of magnetism. Skyrmions are topologically nontrivial magnetic textures, they can naturally couple to the magnetic field through Zeeman coupling, and to the itinerant electrons' spin by the Hund's coupling. The former leads to the ac magnetic field-induced excitation of the Skyrmion internal modes \cite{mochizuki,onose,tchernyshyov} and the latter to the electric current-induced Skyrmion motion \cite{torque,pfleiderer3,rosch,zang,liu13a,lin13} as well as the counter-effect, the topological Hall effect \cite{pfleiderer3,han1,pfleiderer4}.

In addition to the chiral magnetic metals, a Skyrmion crystal phase was recently discovered in the multiferroic insulator Cu$_2$OSeO$_3$ \cite{Cu2OSeO3,seki12a,seki12b,white}, which opens up the opportunity to study electric field control of the Skyrmions and the related magnetic phases. In this material, the finite electric dipole moment induced for each Skyrmion core can be put to use to control the drift motion of the Skyrmion core by coupling to the electric field gradient \cite{seki12b,liu13b}. Through the inherent magneto-electric coupling in an multiferroic insulator, one can envisage the potential to manipulate the magnetization of the Skyrmions by means of electric field, rather than magnetic field, which is also technologically interesting since it is easier to apply local electric fields using electrodes than it would be to create local magnetic fields.

A first step towards exploiting such properties is a framework to predict and to measure the magneto-electric response, which is not limited to the Skyrmion crystal phase but may manifest itself also in the non-topological helical and conical phases. It means, among others, that different magneto-electric responses of those phases can serve as a probe of the magnetic structure as well as the phase transitions between them. In this paper, we show that the Skyrmion crystal phase exhibits a (quantitatively) different magneto-electric response compared to either conical or helical phases. All the principal phases show linear dependence of the magneto-electric (ME) susceptibility $\chi_\mathrm{ME}$ on the applied magnetic field strength $B$, with different slopes. Our theoretical explanation of the ME response of the multiferroic, helical magnetic insulator is based on extensive Monte Carlo (MC) simulations and Ginzburg-Landau (GL) theory. These two methods are complementary to each other and when both of them are applicable, the results show excellent agreement.

The paper is organized as follows. We show our model Hamiltonian in Sec.~\ref{sec:model}. Extensive MC results for ME responses and other thermodynamic quantities are presented in Sec.~\ref{sec-2D-MC} for two dimensions, followed by similar MC calculations for the three-dimensional lattice system in Sec.~\ref{sec-3D-MC}. Ginzburg-Landau theory for ME susceptibility is developed in Sec.~\ref{sec-ME-in-GL}. Outlook and discussion are given in Sec.~\ref{sec-conclusion}.

\section{Model Hamiltonian} \label{sec:model}

The starting point in our calculation of the magneto-electric susceptibility $\chi_{\rm ME}=\partial M / \partial E$ is the three-dimensional cubic-lattice Hamiltonian $H=H_\mathrm{HDM} + H_\mathrm{ME}$ written in terms of the unit classical spin vector $\v S_i$. The first term $H_\mathrm{HDM}$ describes the Heisenberg-Dzyaloshinskii-Moriya spin exchange Hamiltonian together with the Zeeman field, known to yield the helical-to-Skyrmion crystal phase transition as a function of magnetic field. It reads \cite{mochizuki,han2,liu11,han1}
\ba
H_\mathrm{HDM} &=&
\sum_{\v r, \hat e} (-J\,{\v S_{\v r}}\cdot
{\v S_{\v r+a\hat e}} - D\,\hat e \cdot {\v S_{\v r}}\times
{\v S_{\v r+a\hat e}}) \nn
&&-\sum_{\v r}\v B \cdot \v S_{\v r},
\label{eq:H_HDM}
\ea
where $\hat e = \hat x, \hat y, \hat z$ is the bond vector together with $a$ the lattice spacing on a cubic lattice. The second term, $H_\mathrm{ME}$, is the magneto-electric coupling specific to the multiferroic insulator Cu$_2$OSeO$_3$ where the Skyrmion phase has been recently discovered \cite{Cu2OSeO3}. In deriving $H_\mathrm{ME}$ we assume the existence of local electric dipole moment $\v P_i$, coupled to the external electric field $\v E$ in the usual manner $- \v E \cdot \sum_i \v P_i$. A critical insight pointed out in a number of recent papers \cite{liu13b,ansermet,gong12} is that $\v P_i$ depends on the local magnetization configuration according to
\ba
\v P_i = \lambda (S^y_i S^z_i , S^z_i S^x_i , S^x_i S^y_i )
\label{eq:pd-coupling}
\ea
with some coupling strength $\lambda$. This is the ``$pd$-hybridization'' mechanism of multiferroics satisfying cubic symmetry \cite{seki12b,jia1,jia2,arima,penc,BCGO}. As a result, we obtain the Hamiltonian
\ba
H_\mathrm{ME} = -\sum_i \v P_i \cdot \v E_i = -{\lambda \over 2}
\sum_i \v S_i
\bpm
0 & E_i^z & E_i^y \\
E_i^z & 0 & E_i^x \\
E_i^y & E_i^x & 0
\epm
\v S_i .
\label{eq:H-ME}
\ea
Our model is a coarse-grained version of the detailed crystal structure present in the actual material, so $\v S_i$ and $\v P_i$ are both average quantities inside a crystal unit cell. The magneto-electric response will depend on the direction of both the magnetic field and of the electric field. The magnetic field $\v B$ determines the plane wherein the Skyrmions form (orthogonal to $\v B$). The electric field $\v E$ can be applied along another direction to probe, for example, the magneto-electric response $\partial M/\partial E$, where the magnetization $M$ is the thermodynamic average of spin per site along a measuring direction. We calculate in the following the case of $\v B\parallel$ [111] and $\v E\parallel$ [111], i.e. both $\v B$ and $\v E$ fields oriented along the [111] crystal axis. After a rotation transformation \cite{liu13b} $\mathcal{R}$, which places the new $z$-axis along $[111]$ ($\hat{z}'=\mathcal{R}\hat{z}\parallel[111]$) and the new $x$-axis along $[\bar{1} 1 0]$ ($\hat{x}'=\mathcal{R}\hat{x}\parallel[\bar{1}10]$), the magneto-electric coupling is simplified to a uniaxial anisotropy term $H_\mathrm{ME}=-(\sqrt{3} \lambda E / 2) \sum_i [S_i^{z'}]^2$. In the following discussion we absorb $\lambda$ into $E$ by $E'=\lambda E$, so that $E'$ has dimension of energy. Primes are dropped from here on.

The magneto-electric response function is worked out according to standard rules of statistical mechanics,
\begin{widetext}
\ba
\chi_\mathrm{ME} = {\partial {\v M} \over \partial E} \cdot \hat{a} =
{1 \over NT} \left( \Big\langle \Big[ \sum_i \v S_i \cdot \hat a \Big] \Big[ \sum_i \v P_i \cdot \hat b \Big] \Big\rangle - \Big\langle \sum_i \v S_i \cdot \hat a \Big\rangle
\Big\langle \sum_i \v P_i \cdot \hat b \Big\rangle \right),
\label{eq:MC-XME1}
\ea
\end{widetext}
where $\v M = \langle \sum_i \v S_i \rangle /N$ is the thermodynamic average of magnetic moment per site, $N$ is the number of lattice sites, and $T$ is the temperature. $\hat a$ is the direction along which we measure the magnetic moment, $\hat b$ is the direction of the external electric field. The canonical ensemble averages $\langle \cdots \rangle$ are performed by the method of Monte Carlo (MC) simulation. Unless otherwise noted, the magnetization measured in the magneto-electric susceptibility is assumed parallel to the magnetic field applied. In the case of $\hat{a}\parallel\hat{b}\parallel[111]$, after the rotation, we have
\ba
\sum_i \v S_i \cdot \hat{a} &=& \sum_i S_i^z, \nn
\sum_i \v P_i \cdot \hat{b} &=& -{\partial H_\mathrm{ME} \over \partial E} = {\sqrt{3} \over 2} \sum_i [S_i^z]^2. \label{eq:rot}
\ea

\section{Monte Carlo calculation}

\subsection{Film material} \label{sec-2D-MC}

\begin{figure}[h]
  \includegraphics[width=85mm]{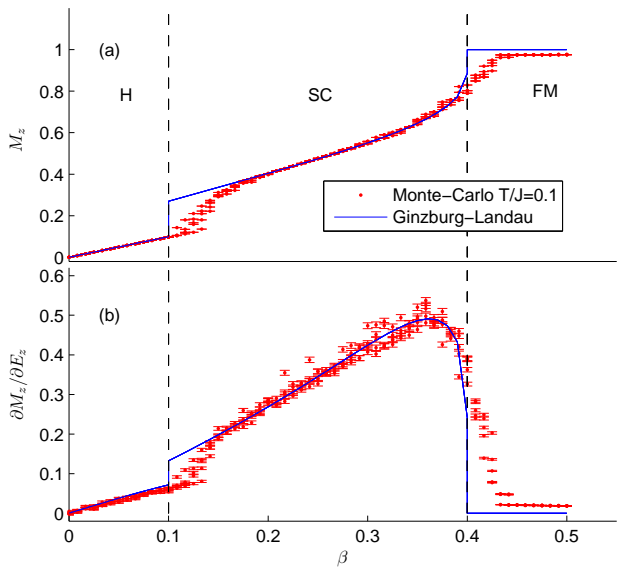}
  \caption{(Color online) (a) Magnetization curve $M_z$ and (b) magneto-electric susceptibility $\chi_\mathrm{ME}$ for varying magnetic field $\beta$ in the 2D helical, Skyrmion crystal and ferromagnetic phases. The electric and magnetic fields are re-scaled to dimensionless values, see   Sec.~\ref{sec-ME-in-GL}.} \label{fig:2dmc}
\end{figure}

\begin{figure}[h]
  \includegraphics[width=85mm]{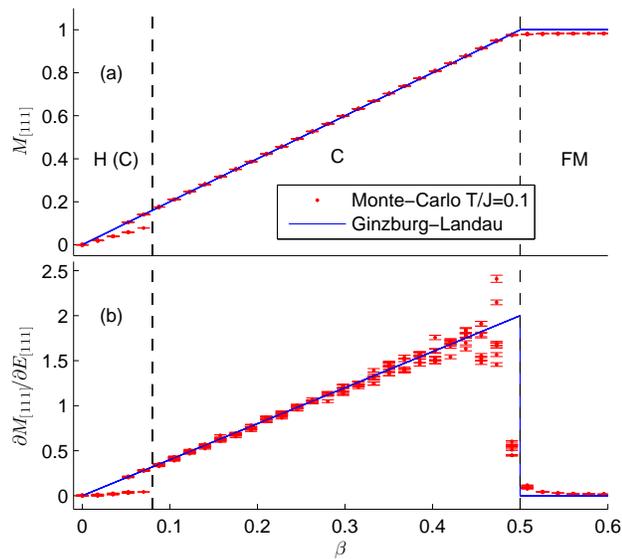}
  \caption{(Color online) (a) Magnetization curve $M_{[111]}$ and (b)   magneto-electric susceptibility $\chi_\mathrm{ME}$ for varying magnetic fields $\beta$ in the 3D conical phase in low temperature $T/J=0.1$. All quantities are re-scaled to dimensionless form, see Sec.~\ref{sec-ME-in-GL}. } \label{fig:3dmc_c}
\end{figure}

It has been found experimentally that for films of thicknesses comparable to a few times the helical modulation period, the Skyrmion lattice phase is stabilized over most of the phase diagram where the corresponding bulk sample displays the conical order \cite{FeGe}. Therefore, the properties of the Skyrmion phase and the transition from the helical to Skyrmion phase can be studied theoretically by simulating a 2D lattice allowing larger system sizes. Here we study the film material perpendicular to the $[1 1 1]$ crystal direction, the electric and magnetic fields are in this direction too. The lattice size is $N=36^2$, and the modulation wavelength is chosen to be 9 lattice constants by taking the ratio $D/J=\sqrt{2}\tan(2\pi/9)$. Figure~\ref{fig:2dmc} shows the $\chi_\mathrm{ME}$ curve as a function of increasing magnetic field $B$ along with the magnetization curve $M(B)$. To make comparison with the Ginzburg-Landau (GL) results feasible, low temperature was used in the simulation. We choose $T/J=0.1$, where $J$ is the exchange energy, comparable to the magnetic ordering transition temperature $T_c$ in the model. Two magnetic phase transitions going from helical (H) to Skyrmion crystal (SC) phase at a lower critical field $B=B_{c1}$, and the other going from SC to ferromagnetic (FM) phase at the upper critical field $B=B_{c2}$, are clearly identified by abrupt changes in the measured quantities in Fig. \ref{fig:2dmc}. Judging from the larger value of $\chi_\mathrm{ME}$ within the Skyrmion phase we conclude that there is more magneto-electric susceptibility per area coming from the Skyrmion phase than the helical phase. Interestingly, both helical and Skyrmion phases exhibit a linear dependence of $\chi_\mathrm{ME}$ on the magnetic field strength $B$, with differing slopes.  We will show how this feature follows from the Ginzburg-Landau analysis in Sec.~\ref{sec-ME-in-GL}.

The helical and Skyrmion crystal phases spontaneously break the spatial translation symmetry, these nonuniform phases host many meta-stable states, so we have plotted in the figure results from multiple runs. We use the ``field-cooling'' method by fixing the magnetic field to the target field and lowering temperature from above $T_c$ down to the target temperature of interest. Within each run an estimation of the statistical error is made by the ``bootstrap method''. The bootstrap error bars are smaller than the difference between results from different runs. Multiple-run results agree with each other reasonably well; this is because the field cooling method gives the right thermodynamic state in this case. The results from GL and MC show some minor differences. The MC curve of magnetization is not saturated in the high field region due to the finite temperature effect. Because of the finite size effect, the transition magnetic field produced by MC is shifted from the GL result, and the MC phase transition is not as sharp as GL because there is phase coexistence in MC, which is absent in GL.

\subsection{Bulk Material}
\label{sec-3D-MC}

\subsubsection{Conical phase at low temperature}

Prior simulations of the multiferroic chiral magnet has been limited to two-dimensional lattice. The 2D simulation of the previous section misses the conical (C) phase and therefore makes the direct comparison to the ME susceptibility experiments, done on a three-dimensional single crystal, difficult. To discuss the magneto-electric response of a 3D lattice in the conical phase, we perform a Monte Carlo simulation dedicated to the conical phase on a $N=10^3$ lattice with helical wavelength equal to 10 lattice constants, i.e., $D/J=\sqrt{3}\tan(2\pi/10)$. The electric and magnetic fields are both put along $[111]$, as before. With these parameters, the system favors conical phase along $[111]$, as easily verified by directly observing the typical configurations generated by the Monte Carlo process. To measure the magneto-electric susceptibility, we calculated the correlation function between $S_{[111]}=\sum_i(S_i^x+S_i^y+S_i^z)/\sqrt{3}$ and $P_{[111]}=\sum_i(P_i^x+P_i^y+P_i^z)/\sqrt{3}$, which reads
\ba
\chi_\mathrm{ME} = {1 \over {NT}}
\Big(\left<S_{[111]}P_{[111]}\right>-
\left<S_{[111]}\right>\left<P_{[111]}\right>\Big).
\ea
No rotation transformation is needed in this simulation. The temperature is chosen again to be $T/J=0.1$. Results are shown in Fig.~\ref{fig:3dmc_c}. In the low field region, besides the conical phase with the modulation vector parallel to $[111]$, helical phases with modulation vector parallel to other $[111]$-equivalent directions also appear. The reason is that the magnetic field is still not large enough to overcome the free energy barrier between helical states with different modulation vectors. So we may conclude that in real bulk material, the system may break into domains with different modulation vectors each pointing to one of the $[111]$ equivalent directions. For large enough magnetic field, only conical phase along $[111]$ is stable, and the magnetization curve is linear with slope equal to two in our dimensionless unit system. This is precisely the result coming from the GL theory (Sec.~\ref{sec-ME-in-GL}). Just like the helical and Skyrmion crystal phases, the magneto-electric susceptibility in conical phase again shows linear behavior with respect to the magnetic field, but with the largest slope 4. The linear behavior and the value of the slope could all be obtained by the GL theory (Sec.~\ref{sec-ME-in-GL}).

In the above discussions, we have studied the magneto-electric response of the chiral magnet in the three principal phases separately. In each of the 2D helical, 2D Skyrmion crystal and 3D conical phases, the magneto-electric susceptibility shows linear dependence on the magnetic field, but with different slopes. The slope of the conical phase is the largest while that of the helical phase the smallest.

\begin{figure*}[ht]
  \includegraphics[width=180mm]{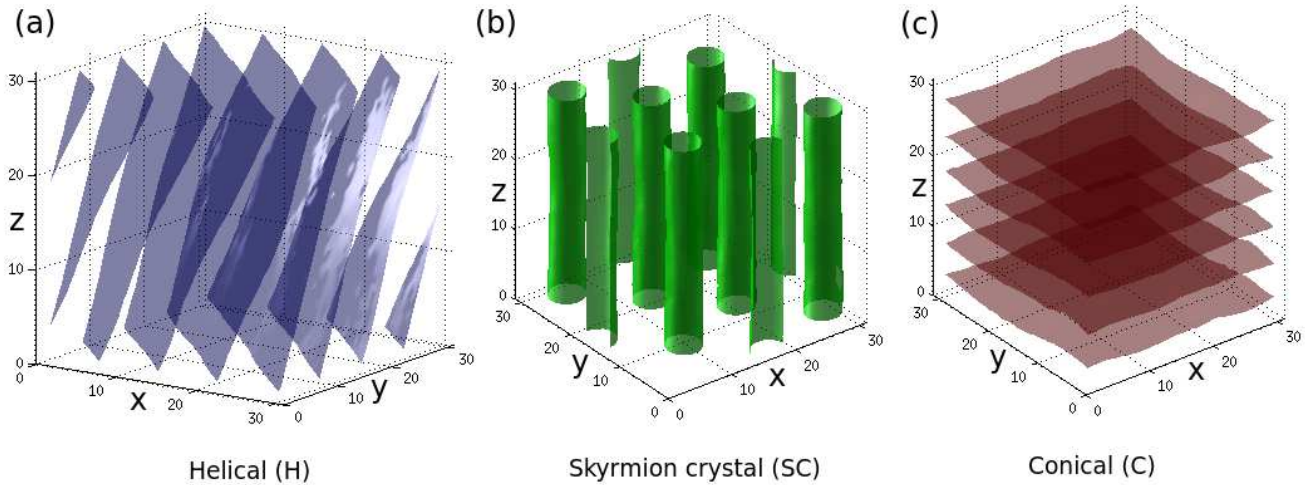}
  \caption{(Color online) Average configuration (order parameter) of the three  dimensional helical (a), Skyrmion crystal (b), and conical (c) phases. The  surfaces of constant value for $S^y=0$, $S^z=0$, and $S^y=0$ are plotted for  the three phases, respectively. In three dimensional helical phase, the  modulation vector is not strictly perpendicular to the magnetic field in  $z$-direction, as opposed to the two dimensional case. The simulation  parameters are: $T/J=0.820$; (a) $B/J=0.025$, (b) $B/J=0.113$, and (c)  $B/J=0.250$.} \label{fig:3dmc_op}
\end{figure*}

\begin{figure}[ht]
  \includegraphics[width=85mm]{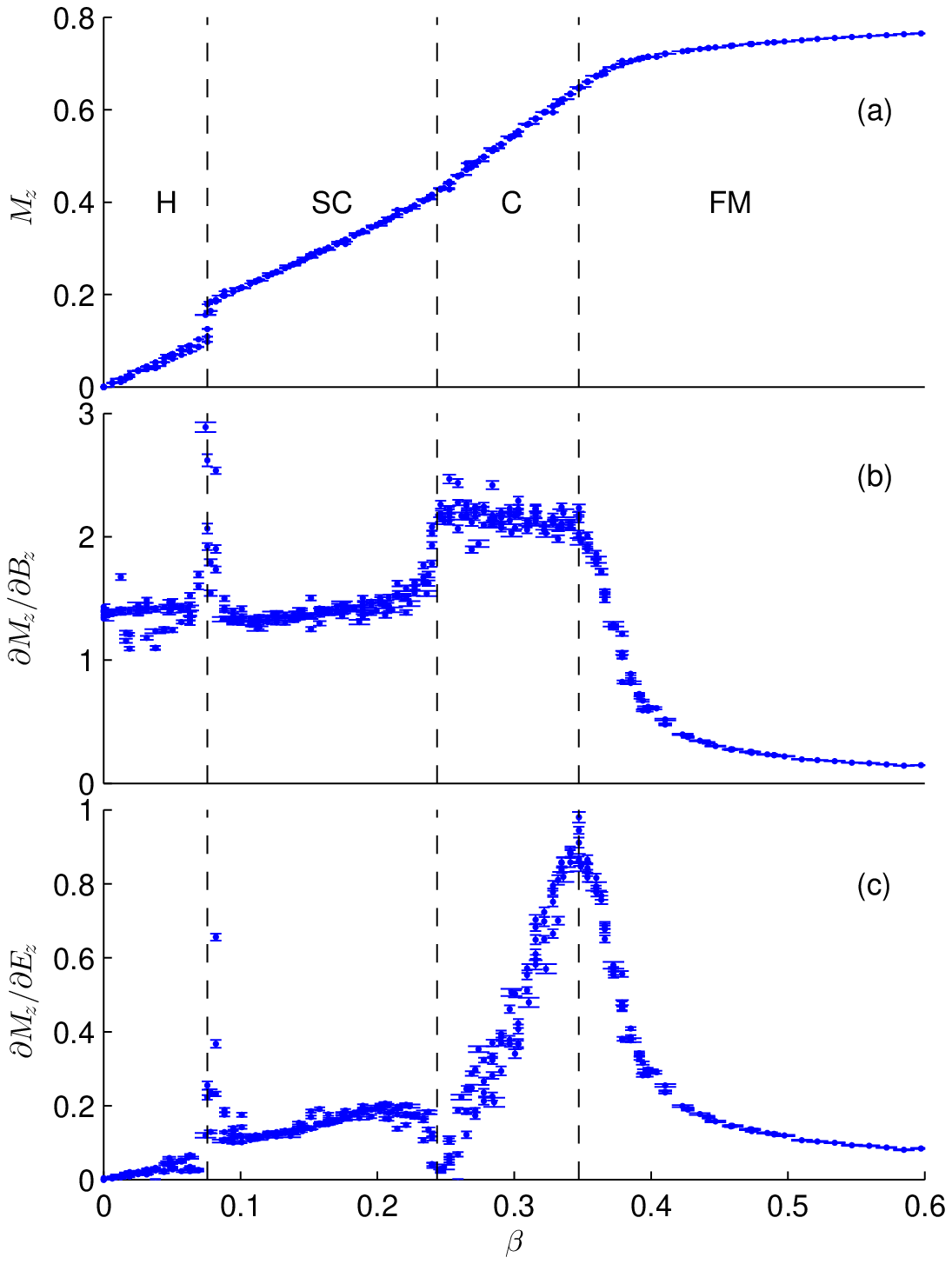}
  \caption{(Color online) (a) Magnetization curve $M_z$ and the magnetic (b) and magneto-electric (c) susceptibilities of a $N=30^3$ lattice with varying  magnetic field $\beta$, the temperature $T/J=0.82$ is just below $T_c/J=0.92$. The helical to Skyrmion crystal and Skyrmion crystal to conical transitions are very easily identified by the peaks and dips of the $\chi_\mathrm{ME}$ curve.  All quantities are re-scaled to dimensionless form, see Sec.~\ref{sec-ME-in-GL}.} \label{fig:3dmc_all}
\end{figure}

\subsubsection{3D simulation just below $T_c$}

Having obtained the magneto-electric response of all the relevant phases separately, it is now desirable to perform a simulation showing all the phases and compare with previous results. Very recently, Buhrandt and Fritz successfully obtained the 3D A-phase of chiral magnet just below the ordering temperature in a Monte Carlo simulation \cite{fritz}. The key insight of their work is to remove the lattice induced anisotropy by introducing next-to-nearest neighbor coupling constants. In this way, the conical phase becomes the thermodynamically stable phase over a lower temperature range and the so-called A-phase of Skyrmion crystal is only stable in a narrow window just below $T_c$. It is to be noted that the lattice-induced anisotropy is only important when the Skyrmion crystal phase and conical phase compete, so it does not affect the previous results we have obtained for the 2D system and the 3D conical phase.

To study the finite temperature (just below $T_c$) magneto-electric response, we adopt the Buhrandt-Fritz model with the following parameters: $N=30^3$, $D/J=\tan(2\pi/10)$, $J'/J=-1/16$, $D'/D=-1/8$, and $T/J=0.82$. Here $J'$ and $D'$ are the Heisenberg and DM interactions between ${\v S}_{\v r}$ and ${\v   S}_{\v r + 2 a \hat e}$. $T_c$ in zero magnetic field is around $0.92J$. Typical magnetic orders are shown in Fig.~\ref{fig:3dmc_op}. The results for the magnetic and magneto-electric susceptibilities are shown in Fig.~\ref{fig:3dmc_all}. In a 3D system as large as $N=30^3$, the number of competing meta-stable states are much larger than before. Luckily, multiple runs of the Monte Carlo sampler give qualitatively the same result. The electric and magnetic fields are directed to $z$-direction of the lattice, but we still map this direction of the simulated lattice to the $[111]$ direction of the modeled bulk material. So to calculate the magneto-electric susceptibility, we still use the formula Eq.~(\ref{eq:MC-XME1}) and Eq.~(\ref{eq:rot}).

From the magnetization curve (Fig.~\ref{fig:3dmc_all}(a)) we could easily see the helical to Skyrmion crystal transition by the abrupt increase of total magnetic moment in $z$-direction. The Skyrmion crystal to conical phase transition is identified by the increase of magnetic susceptibility (Fig.~\ref{fig:3dmc_all}(b)). Overall we find the magneto-electric susceptibility curve gives much more clear boundary between the different phases by the peaks and dips, compared to more conventional magnetization or magnetic susceptibility curves. The magneto-electric response is still largest in the conical phase and smallest in the helical phase.

\section{Magneto-electric susceptibility in Ginzburg-Landau theory}
\label{sec-ME-in-GL}

As stated above, one interesting new possibility raised by the multiferroic chiral magnetic material is that of measuring the magneto-electric response in the Skyrmion crystal phase. Even the non-topological states, such as helical or conical phases, provide a non-trivial ME response. In this section we provide Ginzburg-Landau calculation of this quantity, which accurately captures the ME responses obtained from the previous MC simulations at low temperatures. The GL analysis just below $T_c$ is more complicated and will be addressed in the following works.

We use the standard chiral magnet free energy with Heisenberg and DM exchanges and the Zeeman term. The magneto-electric coupling is expressed by
\ba
F_\mathrm{ME}=-(E_xM_yM_z+\mathrm{cyclic\ perm.}).
\ea
We again choose the $\v B \parallel \v E \parallel [111]$ geometry under which experiments are often carried out. The GL expressions simplify greatly in this setup as well. The full GL free energy takes the form
\ba F & = &
\frac{J}{2}\left(\nabla\mathbf{M}\right)^{2}+D\mathbf{M}\cdot
\left(\nabla\times\mathbf{M}\right)\nn
&&-B\left(M_{x}+M_{y}+M_{z}\right)/\sqrt{3}\nn
&&-E\left(M_{y}M_{z}+M_{z}M_{x}+M_{x}M_{y}\right)/\sqrt{3},
\label{free-eng-xyz}
\ea
which could be obtained by the standard continuum limit process in three dimensions from the lattice Hamiltonian Eq.~(\ref{eq:H_HDM}) with the following relations
\ba
J \rightarrow a J, \quad D \rightarrow a^2 D, \quad
B \rightarrow a^3 B, \quad E \rightarrow a^3 E.
\ea
Then we perform a joint real and spin space rotation
\ba
({\cal R}{\v M})(\v r)={\cal R}({\v M}({\cal R}^{-1}{\v r})) \nonumber
\ea
as before, which makes the new $z$-axis along the $[111]$-direction and the new $x$-axis along the $[\bar{1} 1 0]$-direction \cite{liu13b}, so the free energy becomes
\ba
{F \over 8J \kappa^2} =
\left(\nabla\mathbf{M}\right)^{2}+\mathbf{M}\cdot
\left(\nabla\times\mathbf{M}\right) -\beta M_z-{1\over2}\epsilon M_z^2.
\label{free-energy-xyz-rotated}
\ea
Note the dimensionless magnetic and electric fields $\beta=B/(8J\kappa^2)$ and $\epsilon=\sqrt{3}E/(8J\kappa^2)$ in the above. The space coordinates are re-scaled as well, $\v r \rightarrow \v r/(4\kappa)$, where $\kappa = D/(2J)$ is half of the magnitude of the helical modulation vector. The dimensionless free energy in Eq.~(\ref{free-energy-xyz-rotated}) is the starting point of the following discussion.

\subsection{Helical phase}
First we consider the helical phase in the thin film geometry with the magnetic field perpendicular to it. For the fields not exceeding the threshold value $\beta_{c1}$ the system is in helical phase. It can be proven analytically that without magnetic field, the ground state is the so-called proper-screw state
\ba
{\v M}(x,y,z)=(0,\cos(x/2),\sin(x/2)).
\ea
Here we have chosen the spin modulation vector ${\v q}\parallel\hat{x}$ as an example. In fact the free energy (\ref{free-energy-xyz-rotated}) is isotropic in the whole space so ${\v q}$ could point to any direction as long as the magnitude is unchanged.

\begin{figure}[t]
  \includegraphics[width=85mm]{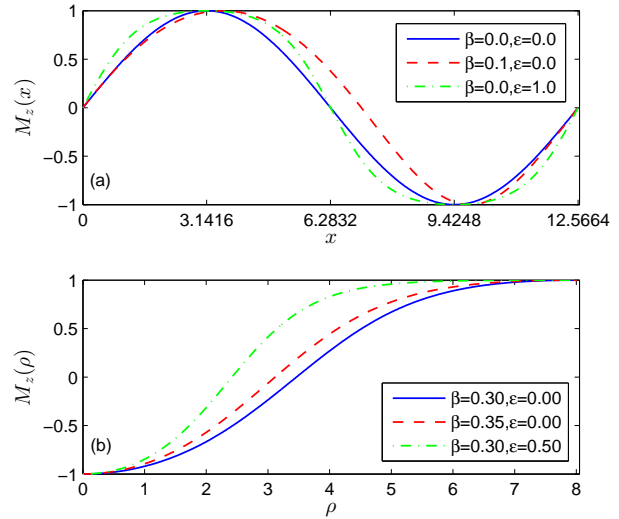}
  \caption{ (Color online) The spatial profile of the $z$-component of  magnetization in different electric and magnetic fields for the helical (a)  and Skyrmion crystal (b) phases. Larger magnetic field favors more  magnetization pointing up, while larger electric field favors more  magnetization pointing up or pointing down. } \label{fig:twist}
\end{figure}

Now we turn on the electric and magnetic fields. Monte Carlo results show that as long as neither fields are very big, the ground state is still a proper screw, but the modulation vector is not constant, so we make the following ansatz to characterize this twisted screw
\ba
{\v M}(x,y,z)=(0,\cos(\theta(x)),\sin(\theta(x))).
\ea
Now $\theta(x)$ becomes a variational function to be optimized with respect to Eq.~(\ref{free-energy-xyz-rotated}). At finite fields, the optimized function deviates from the zero field solution $\theta(x)=x/2$. After inserting this state to the free energy we have
\ba
{F[\theta(x)] \over 8J\kappa^2}&=&[\theta'(x)]^2-\theta'(x) \nn
&&-\beta \sin(\theta(x))-{1\over2}\epsilon \sin^2(\theta(x)).
\ea
This functional is to be optimized with boundary conditions $\theta(0)=0$ and $\theta(4\pi)=2\pi$, which is compatible with the wavelength and the right-handedness of the proper screw state in the absence of the fields. The numerical calculation is done as follows. We sweep magnetic field $\beta$ from 0 to 0.1, with 0.1 being the lower critical field from helical phase to Skyrmion crystal phase \cite{bogdanov1,han2,liu13a} transition. At each magnetic field, we calculate two optimized functions $\theta(x)$. One with $\epsilon=0$ and one with $\epsilon=\delta\epsilon$. Finally, the magneto-electric susceptibility is determined from
\ba
{\partial M_z \over \partial \epsilon}&=&\int_0^{4\pi}dx\times\nn
&&{\sin[\theta(x;\epsilon=\delta \epsilon)]-\sin[\theta(x;\epsilon=0)]\over
4 \pi \times \delta \epsilon}.
\ea
Figure~\ref{fig:2dmc} shows the results obtained in this way with $\delta \epsilon = 0.05$, which agree with the MC very well. We have checked that $\delta \epsilon = 0.05$ is small enough so that the system is in the linear response region. In Fig.~\ref{fig:twist}(a) we plot three typical functions $\sin(\theta(x))$, which is the $z$-component of the local magnetization, in different fields. The blue curve shows a perfect proper screw state with $\theta(x)=x/2$. The red curve is the twisted screw in finite magnetic field and zero electric field. It is clear that the region with magnetization pointing up is enlarged compared with the perfect screw, since the Zeeman term favors magnetization pointing in that direction. The green curve shows the twisted screw in finite electric field and zero magnetic field, in which both the regions with magnetization pointing up and pointing down are enlarged. This is because the electric field induces an easy $z$-axis anisotropy.

\subsection{Conical phase}

Now we consider the conical phase in a 3D material. The right-handed conical spin configuration which is compatible with $D>0$ follows
\ba
{\v M}(x,y,z)=
(\sin(\theta)\cos(qz),\sin(\theta)\sin(qz),\cos(\theta))
\ea
where $\theta$ is the conical angle, $(0,0,q)$ is the conical modulation vector. Insert it into the energy functional Eq.~(\ref{free-energy-xyz-rotated}), we have the energy density as a function of just two variables $q$ and $\theta$
\ba
{F(q,\theta)\over 8J\kappa^2}\!=\!{1\over 2} (q^2 \!-\!
q)(1\!-\!\cos(2\theta) ) \!-\!\beta \cos(\theta) \!-\!
{1\over2}\epsilon\cos^2(\theta). \nonumber
\ea
Minimizing this function with respect to $q$ and $\theta$ we get
\ba
q_{0} = {1\over2}, ~~ \cos(\theta_{0}) = {2\beta\over 1-2\epsilon}.
\ea
It shows that the magnetic and electric fields in $[111]$-direction does not change the modulation vector. The magnetic susceptibility is
\ba
\frac{\partial M_z}{\partial \beta}=
{2 \over 1 -2\epsilon},
\ea
which does not depend on the magnetic field but depends on the electric field. The magneto-electric susceptibility is
\ba
\frac{\partial M_z}{\partial \epsilon}=
{4 \beta \over (1-2\epsilon)^2},
\ea
which depends on the magnetic field linearly. The value at $\epsilon=0$, in particular is $4\beta$. These conclusions agree with the MC results very well, see Fig.~\ref{fig:3dmc_c}.

\subsection{Skyrmion crystal phase}

Finally we consider the Skyrmion crystal phase which appears in both 2D and 3D materials. Following the method of previous works
\cite{bogdanov1,bogdanov2,liu13a,han2}, we start from the trial function of a single Skyrmion
\ba \mathbf{M}\left(\rho,\phi,z\right)=
\sin\left[\theta\left(\rho\right)\right]\hat{\phi}+
\cos\left[\theta\left(\rho\right)\right]\hat{ z},
\ea
where $\theta(\rho)$ is the variational function to be optimized with respect to Eq.~(\ref{free-energy-xyz-rotated}), which, after inserting the trial function, becomes
\ba
\frac{F\left(\rho\right)}{8J\kappa^2} & = &
\left(\frac{\partial\theta}
{\partial\rho}\right)^2+\frac{\partial\theta}{\partial\rho}
+{\sin^2(\theta) \over \rho^2}+{\sin(\theta)\cos(\theta)\over \rho}\nn
&&-\beta\cos(\theta)-{1\over2}\epsilon\cos^2(\theta).
\ea
It is already known that the phase boundary between the helical and Skyrmion crystal phases is located at $\beta_{c1}=0.1$ and that between the Skyrmion crystal and ferromagnetic phases is located at $\beta_{c2}=0.4$ \cite{bogdanov1,han2,liu13a}. To analyze the Skyrmion crystal, we need to introduce a radius cut $R_{0}$, which gives the region occupied by each Skyrmion in the lattice. Then we optimize $\theta\left(\rho\right)$ with the boundary conditions $\theta\left(0\right)=\pi$ and $\theta\left(R_{0}\right)=0$. Finally we calculate the average energy density
\ba
{ \int_{0}^{R_{0}} d\rho ~ 2 \pi \rho \, F (\rho )/ (8J\kappa^2 )
  \over \pi R_{0}^{2} }\nonumber
\ea
and search for $R_{0}$ so that the average energy density is optimized. This corresponds to finding the optimized lattice spacing for the Skyrmion lattice. The result $R_0$ depends on the magnetic field $\beta$ \cite{bogdanov1,han2}.

With the above background, we again sweep the magnet field $\beta$, but this time from 0.1 to 0.4. At each $\beta$ we calculate two optimized functions $\theta\left(\rho\right)$, one with $\epsilon=0$ and one with $\epsilon=\delta \epsilon$. These two functions both use the cut $R_0$ for this magnetic field. Then we calculate the magneto-electric susceptibility
\ba
\frac{dM_z}{d\epsilon}&=&\int_{0}^{R_{0}}2\pi\rho \, d\rho \, \times \nn
&&\frac{
\cos\left[\theta\left(\rho;\epsilon=\delta \epsilon\right)\right]-\cos
\left[\theta\left(\rho;\epsilon=0\right)\right] }
{\pi R_{0}^{2}\times \delta \epsilon}.
\ea
The result with $\delta \epsilon=0.05$ is shown in Fig.~\ref{fig:2dmc}, which has remarkable agreement with MC. In Fig.~\ref{fig:twist}(b) we show the function $\cos(\theta(\rho))$ in different fields. The blue curve is the $z$-component of magnetization of a Skyrmion inside a unit cell of the Skyrmion lattice. The red curve is the result with larger magnetic field, in which the peripheral region with magnetization pointing up is enlarged. The green curve is the result in nonzero electric field, in which the peripheral region is also enlarged. Notice that the central region is not enlarged because in cylindrical coordinate, the central region has weight $2\pi\rho$ which is much smaller than the peripheral region. As a result, the enlargement of the peripheral region gives more energy gain.

\subsection{Estimation}

Figure \ref{fig:2dmc} shows the Ginzburg-Landau together with the Monte Carlo results. In the following the energy unit $J$ is set to 1. In the 2D Monte Carlo simulation $\kappa=D/(2J)=\sqrt{2}\tan(2\pi/9)/2=0.59$. According to our re-scaling, the electric field, magnetic field and energy density all scale as $1/(8J\kappa^2)=0.36$. Since Monte Carlo has finite size effect, the calculated best fit parameter is $1/(8J\kappa^2)=0.42$. The results from the two methods agree quite well except at the phase boundary. The reason, as we have mentioned, is that in Ginzburg-Landau calculation the phase transition is very sharp but in Monte Carlo there is always a phase coexists at the phase boundary. The 3D conical state simulation is re-scaled with simulation parameter $\kappa=D/(2J)=\sqrt{3}\tan(2\pi/10)/2=0.63$. The re-scaling parameter is $1/(8J\kappa^2)=0.32$ but the best fit parameter is $1/(8J\kappa^2)=0.35$.  The results are plotted in Fig.~\ref{fig:3dmc_c}. It is shown that the Ginzburg-Landau result $\partial M / \partial \epsilon = 4 \beta$ agrees very well with the Monte Carlo result.

We can also substitute the material parameters into our dimensionless unit system and give estimations as follows. First we calculate the dimensionless values corresponding to one Tesla of magnetic field and one Volt per nanometer of electric field,
\ba
\beta_0 &=& {4 \mu_B
\times (1\mathrm{T})\over 8J\kappa^2} = 2.94 \nn \epsilon_0 &=&
{\sqrt{3} \lambda \times (1\mathrm{V/nm})\over 8J\kappa^2} = 1.37,
\ea
where $4\mu_B$ is the magnetic moment inside a unit cell of Cu$_2$OSeO$_3$; the exchange strength is taken to be $J=5\mathrm{meV}$; $\kappa=D/(2J)=\pi/l$ with $l=630/8.9$ the wavelength of helix in the unit of lattice constant $a=8.9\mathrm{nm}$; the magneto-electric coupling constant is estimated to be $\lambda=1\times10^{-32}\mathrm{J/(V/m)}$. These set of parameters is taken from Ref.~\onlinecite{seki12b} and the typical ME-susceptibility in conical phase is thus
\ba
{\partial M \over \partial E} = {\partial M \over \partial \epsilon}{\partial
\epsilon \over \partial E} = {4 \beta \epsilon_0 \over \mathrm{(1V/nm)}} =
0.55\mathrm{(V/nm)}^{-1}
\label{eq:estimation}
\ea
with typical value $\beta=0.1$. This result also corresponds to $0.14\mathrm{(\mu_B/Cu)/(V/nm)}$ since the saturate magnetic moment is $4\mu_B$ and there are 16 copper atoms in a unit cell (three up one down structure). Thermal fluctuation generally reduces the average magnetic moment from the ground state value, so the magneto-electric susceptibility in finite temperature should be less than the Ginzburg-Landau estimation, which may lead to minor over-estimation.

\section{Discussion and outlook}
\label{sec-conclusion}

In summary, inspired by the recent discovery of the multiferroic chiral magnetic material Cu$_2$OSeO$_3$, whose underlying mechanism of magneto-electric coupling is the so-called ``$pd$ hybridization'', we have performed theoretical analysis of the magneto-electric response $\partial M / \partial E$ for this class of material. We studied the magneto-electric response in all the relevant magnetic phases. We found that the magneto-electric response in the $\v B \parallel \v E \parallel [111]$ geometry shows linear dependence on the magnetic field in all the helical, conical and Skyrmion crystal phases, but with different slopes. Estimation using realistic material parameters is also given to compare with potential experiments. In addition to the usual magnetic susceptibility, the magneto-electric susceptibility gives a new way to identify the magnetic phases transitions. We also showed how the electric field modulates the magnetic structures, which could be used in future applications.

\begin{acknowledgements}
 Y. H. L. and Y. Q. L. are supported by the NSFC grant (No.~11074216 \&  No.~11274272). J. H. H. is supported by  the NRF grant (No. 2013R1A2A1A01006430). A. A. O. and H. M. R. are supported by the Swiss National Science Foundation and the European Research Council grant CONQUEST. Y. H. L. would like to thank Q. Zhu and S. Buhrandt for discussions about the Monte Carlo method.
\end{acknowledgements}


\begin{thebibliography}{24}

\bibitem{bogdanov1} A. N. Bogdanov and D. A. Yablonskii, Sov. Phys. JETP
  \textbf{68}, 101 (1989); A. Bogdanov and A. Hubert,
  J. Magn. Magn. Mater. \textbf{138}, 255 (1994).

\bibitem{bogdanov2} U. K. Ro{\ss}ler, A. N. Bogdanov, and C. Pfleiderer, Nature
  \textbf{442}, 797 (2006).

\bibitem{han1} S. D. Yi, S. Onoda, N. Nagaosa, and J. H. Han, Phys. Rev. B
  \textbf{80}, 054416 (2009).

\bibitem{pfleiderer1} S. M\"{u}hlbauer, B. Binz, F. Jonietz, C. Pfleiderer,
  A. Rosch, A. Neubauer, R. Georgii, and P. B\"{o}ni, Science \textbf{323}, 915
  (2009).

\bibitem{pfleiderer2} W. M\"{u}nzer, A. Neubauer, T. Adams, S. M\"{u}hlbauer,
  C. Franz, F. Jonietz, R. Georgii, P. B\"{o}ni, B. Pedersen, M. Schmidt,
  A. Rosch, and C. Pfleiderer, Phys. Rev. B \textbf{81}, 041203(R) (2010).

\bibitem{tokura} X. Z. Yu, Y. Onose, N. Kanazawa, J. H. Park, J. H. Han,
  Y. Matsui, N. Nagaosa, and Y. Tokura, Nature \textbf{465}, 901 (2010).

\bibitem{FeGe} X. Z. Yu, N. Kanazawa, Y. Onose, K. Kimoto, W. Z. Zhang,
  S. Ishiwata, Y. Matsui, and Y. Tokura, Nature Mat. \textbf{10}, 106 (2011).

\bibitem{MnGe} N. Kanazawa, Y. Onose, T. Arima, D. Okuyama, K. Ohoyama,
  S. Wakimoto, K. Kakurai, S. Ishiwata, and Y. Tokura, Phys. Rev. Lett.
  \textbf{106}, 156603 (2011).

\bibitem{mochizuki} M. Mochizuki, Phys. Rev. Lett. \textbf{108}, 017601 (2012).

\bibitem{onose} Y. Onose, Y. Okamura, S. Seki, S. Ishiwata, and Y. Tokura,
  Phys. Rev. Lett. \textbf{109}, 037603 (2012).

\bibitem{tchernyshyov} O. Petrova and O. Tchernyshyov, Phys. Rev. B \textbf{84},
  214433 (2011); I. Makhfudz, B. Kr\"{u}ger, and O. Tchernyshyov,
  Phys. Rev. Lett. \textbf{109}, 217201 (2012).

\bibitem{torque} F. Jonietz, S. M\"{u}hlbauer, C. Pfleiderer, A. Neubauer,
  W. M\"{u}nzer, A. Bauer, T. Adams, R. Georgii, P. B\"{o}ni, R. A. Duine,
  K. Everschor, M. Garst, and A. Rosch, Science \textbf{330}, 1648 (2010).

\bibitem{pfleiderer3} T. Schulz, R. Ritz, A. Bauer, M. Halder, M.Wagner,
  C. Franz, C. Pfleiderer, K. Everschor, M. Garst, and A. Rosch, Nat.
  Phys. \textbf{8}, 301 (2012).

\bibitem{rosch} K. Everschor, M. Garst, R. A. Duine, and A. Rosch, Phys. Rev. B
  \textbf{84}, 064401 (2011).

\bibitem{zang} J. Zang, M. Mostovoy, J. H. Han, and N. Nagaosa,
  Phys. Rev. Lett. \textbf{107}, 136804 (2011).

\bibitem{liu13a} Y. H. Liu and Y. Q. Li, J. Phys.: Condens. Matter \textbf{25},
  076005 (2013).

\bibitem{lin13} S. Z. Lin, C. Reichhardt, C. D. Batista, and A. Saxena,
  Phys. Rev. B \textbf{87}, 214419 (2013).


\bibitem{pfleiderer4} R. Ritz, M. Halder, M. Wagner, C. Franz, A. Bauer, and
  C. Pfleiderer, Nature \textbf{497}, 231 (2013).

\bibitem{Cu2OSeO3} S. Seki, X. Z. Yu, S. Ishiwata, and Y. Tokura,
  \textit{Science} \textbf{336}, 198 (2012).

\bibitem{seki12a} S. Seki, J. H. Kim, D. S. Inosov, R. Georgii, B. Keimer,
  S. Ishiwata, and Y. Tokura, Phys. Rev. B \textbf{85}, 220406(R) (2012).

\bibitem{seki12b} S. Seki, S. Ishiwata, and Y. Tokura, Phys. Rev. B \textbf{86},
  060403(R) (2012).

\bibitem{white} J. S. White, I. Levati\'{c}, A. A. Omrani, N. Egetenmeyer,
  K. Pr\u{s}a, I. \u{Z}ivkovi\'{c}, J. L. Gavilano, J. Kohlbrecher,
  M. Bartkowiak, H. Berger, and H. M. R{\o}nnow, J. Phys.: Condens. Matter
  \textbf{24}, 432201 (2012).

\bibitem{liu13b} Y. H. Liu, Y. Q. Li, and J. H. Han, Phys. Rev. B \textbf{87},
  100402(R) (2013).

\bibitem{han2} J. H. Han, J. Zang, Z. Yang, J. H. Park, and N. Nagaosa,
  Phys. Rev. B \textbf{82}, 094429 (2010).

\bibitem{liu11} Y. Q. Li, Y. H. Liu, and Y. Zhou, Phys. Rev. B \textbf{84},
  205123 (2011).

\bibitem{ansermet} M. Belesi, I. Rousochatzakis, M. Abid, U. K. R\"{o}{\ss}ler,
  H. Berger, and J.-Ph. Ansermet, Phys. Rev. B \textbf{85}, 224413 (2012).

\bibitem{gong12} J. H. Yang, Z. L. Li, X. Z. Lu, M. H. Whangbo, S. H. Wei,
  X. G. Gong, and H. J. Xiang, Phys. Rev. Lett. \textbf{109}, 107203 (2012).

\bibitem{jia1} C. Jia, S. Onoda, N. Nagaosa, and J. H. Han, Phys. Rev. B
  \textbf{74}, 224444 (2006).

\bibitem{jia2} C. Jia, S. Onoda, N. Nagaosa, and J. H. Han, Phys. Rev. B
  \textbf{76}, 144424 (2007).

\bibitem{arima} T. Arima, J. Phys. Soc. Jpn. \textbf{76}, 073702 (2008).

\bibitem{penc} J. Romh\'{a}nyi, M. Lajk\'{o}, and K. Penc, Phys. Rev. B
  \textbf{84}, 224419 (2011).

\bibitem{BCGO} H. Murakawa, Y. Onose, S. Miyahara, N. Furukawa, and Y. Tokura,
  Phys. Rev. Lett. \textbf{105}, 137202 (2010).

\bibitem{fritz} S. Buhrandt and L. Fritz, arXiv:1304.6508 (2013).

\end{thebibliography}
\end{document}